\documentclass{PoS}

\title{The Data Processor of the EUSO-SPB2 Telescopes	}

\ShortTitle{The DP of EUSO-SPB2}

\author{\speaker{Giuseppe Osteria}\\
	Istituto Nazionale di Fisica Nucleare - Sezione di Napoli, Italy
		E-mail: \email{osteria@na.infn.it}}
\author{Francesco Perfetto\\
		Istituto Nazionale di Fisica Nucleare - Sezione di Napoli, Italy}
\author{Valentina Scotti\\
		Universit\`{a} degli Studi di Napoli Federico II, Dipartimento di Fisica, Napoli, Italy and
        Istituto Nazionale di Fisica Nucleare - Sezione di Napoli, Italy\\}
        
\author{for the JEM-EUSO Collaboration \footnote{for collaboration list see PoS(ICRC2019)1177}}

\abstract{In this paper we present the Data Processor (DP) of EUSO-SPB2 (Extreme Universe Space Observatory on a Super Pressure Balloon, mission two) telescopes . The EUSO-SPB2 is the continuation of the JEM-EUSO science program on ultra-long duration balloons, started with the EUSO-SPB1 mission.

The EUSO-SPB2 will host on-board two telescopes. One is a fluorescence telescope designed to detect high energy cosmic rays via the UV fluorescence emission of the showers in the atmosphere; the other one measures direct Cherenkov light emission from lower energy cosmic rays and other optical backgrounds for cosmogenic tau neutrino detection.
The DP is the component of the electronics system which performs data management and instrument control for each of the two telescopes.\

The DP controls front-end electronics, tags events with arrival time and payload position through a GPS system, provides signals for time synchronization of the event and measures live and dead time of the telescope. Furthermore it manages mass memory for data storage and performs housekeeping monitor and controls the power on and power off sequences. Since a super pressure balloon may remain airborne up to 100 days, the requirements on the electronics and data handling are quite severe. The DP operates at high altitude in unpressurised environment which represents a technological challenge for heat dissipation. In this paper we describe the main components of the system and the design developed for the new mission.}

\FullConference{36th International Cosmic Ray Conference -ICRC2019-\\
		July 24th - August 1st, 2019\\
		Madison, WI, U.S.A.}

\begin{document}

\section{Introduction}
The EUSO-SPB2 experiment is the second-generation mission of the JEM-EUSO program \cite{Bertaina}. Its main
objectives are the detection of   Ultra-High Energy Cosmic Rays (UHECR) by measuring fluorescence and Cerenkov light produced by the interaction of the particles with the nuclei of the Earth's atmosphere.
Thereby, EUSO-SPB2 will study for the first time from near space:
\begin{itemize}
	
	\item Cherenkov light from EAS;
	\item the background light from decays of tau leptons induced by up-going tau neutrinos;
	\item fluorescence light from high altitude horizontal showers in a nearly constant density atmosphere for verification of hadronic models at ultra-high energies.
\end{itemize}
More detailed information on the science goals of EUSO-SPB2 can be found in \cite{Adams1}.

EUSO-SPB2 will build upon the experience gained by EUSO-SPB1 flight  \cite{scotti0} in April of 2017.

A number of upgrades, including a Schmidt design telescope and a faster ultraviolet (UV) camera for increased exposure to UHECR observations, will significantly improve the capabilities of EUSO-SPB2. 
In addition, EUSO-SPB2 will serve as a pathfinder for the more ambitious space-based measurements by the Probe Of Extreme Multi-Messenger Astrophysics (POEMMA) \cite{poemma}, currently under study by NASA.
EUSO-SPB2 will be launched from Wanaka (New Zealand) in 2022. The mission aims at a  100 days long stratospheric flight.
A full description of the mission is given in \cite{lawrence}.

\section{The instrument}
To meet the science objectives, two independent telescopes have been designed for specific measurements. Since the characteristics of Cherenkov and fluorescence light are different, two distinctive telescopes are required: a large
Field of View  telescope will observe short Cherenkov signal from horizontal events and from air showers in UV/VIS wavelength range. A second telescope will measure the longer lasting  UV fluorescence signals produced by EAS from UHECR, imaging the shower trajectory as the cascade develops down the atmosphere. 

\begin{figure} 
	\centering
	\includegraphics[width=0.80	\linewidth]{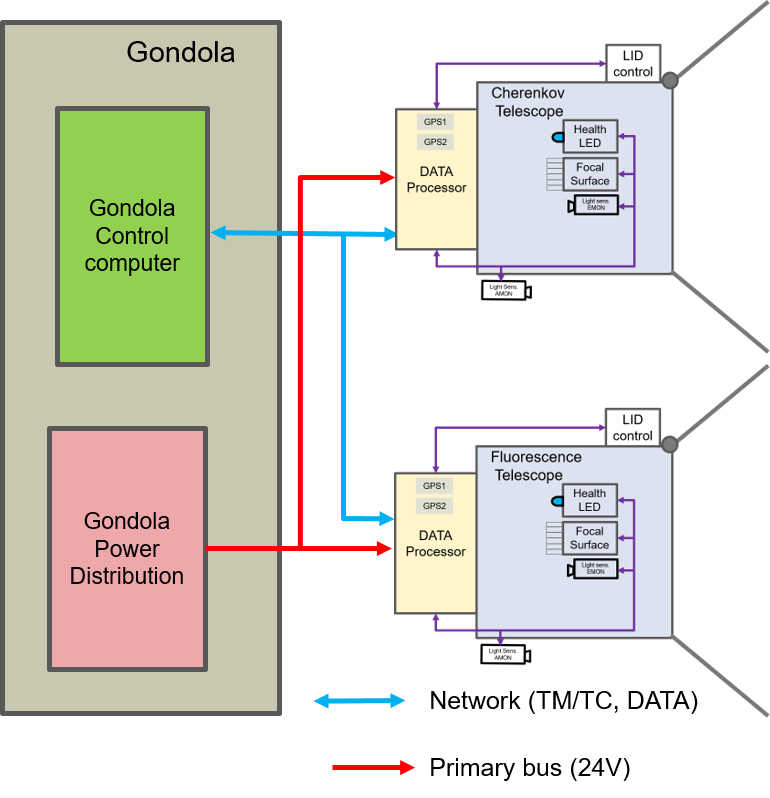}
	\caption{Block diagram of the full system with interfaces between the SIP in the Gondola and the two telescopes. }
	\label{fig:couple_teles}
\end{figure}

Both the telescopes designs consist of a Schmidt optical system, a focal plane detector with associated readout electronics  and a Data Processor (DP). Each imaging systems, has  position resolution of 3 mm for the sensors, and requires a angular resolution of  0.2$ ^{\circ}$. 
The concave focal surface of the fluorescence telescope is composed of three Photon Detection Modules (PDM), each composed of 36 Multi-Anode Photomultiplier Tubes (MAPMT) for almost 7000 pixels in total. Signals from the MAPMTs are processed and digitized with SPACIROC3 ASICs. The SPACIROCs signals are passed to a Zynq board (an FPGA board utilizing the Xilinx Zynq  XC7Z030 chip) which performs the low-level data handling and implements the multi-level trigger scheme \cite{trigger}. 
The Cherenkov telescope focal surface will be covered by silicon photomultipliers (SiPM). They are coupled to 
100 mega-samples per second switch capacitor array (SCA) of the AGET chip (ASIC for General Electronics for TPCs)
An accurate description of the Cherenkov telescope is given in \cite{otte}.
Even though the two telescopes have very different focal surfaces and readout electronics, they 
are connected to identical hardware and sensors as shown in Figure \ref{fig:couple_teles}
Each telescope is equipped with  GPS receivers (for redundancy) used to tag events with absolute time and position. 
To protect the very sensitive focal sensors from sunlight, each telescope is equipped with  a motor controlled lid, which remains closed during the day and opens before starting the data acquisition during the night.
Two light sensors to measure the intensity of the light.
The first device (AMON) \cite{amon} is located outside the telescope. It  measures the  absolute light  intensity in the UV/VIS range from the atmosphere and is also used to evaluate the light reaching the telescope (day/dark detector). The second device (EMON) \cite{amon} is located inside the telescope box and measures the light on the focal surface. It is used to verify dark conditions inside the telescopes during operation and  to diagnose any light leak when the cover is closed.
In addition, a Health LED system  per telescope provides a way to illuminate the full focal surface to check the status of every pixel.

A gondola provides a frame for mounting  both telescopes, the instrument subsystems, and equipment for the balloon mission. 
The Support Instrumentation Package (SIP), built by the Colombia Scientific Balloon Facility, houses the power supplies, flight computers, telemetry systems, and the communication systems. 
The power is distributed through a primary 24V bus and data is routed through an Ethernet network.

%As shown in Figure  the only interfaces between the telescopes and the gondola are power and a network link.

\section{The telescope Data Processor}

\begin{figure} 
	\centering
	\includegraphics[width=0.80	\linewidth]{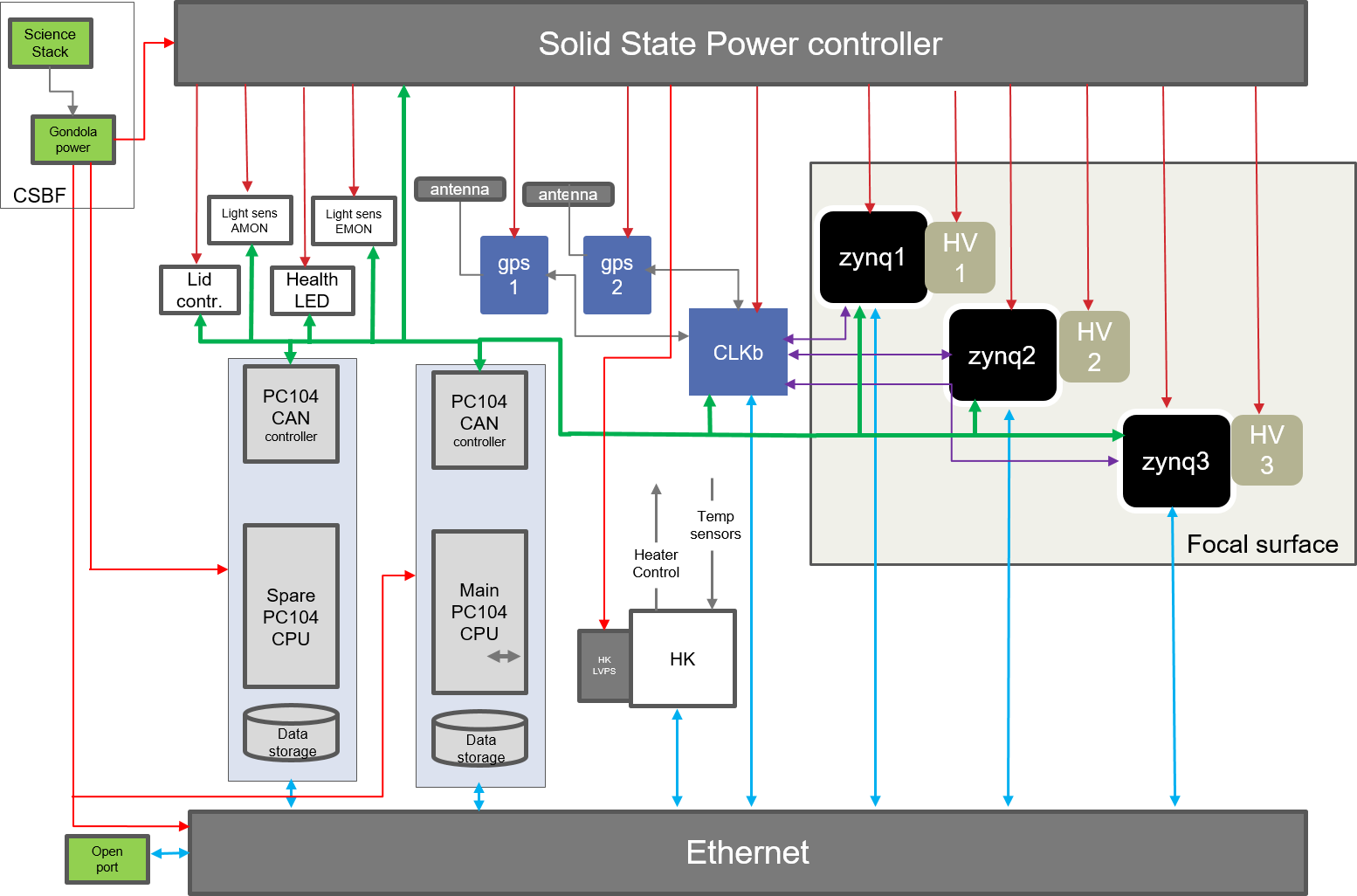}
	\caption{The DP block diagram }
	\label{fig:block_diag1}
\end{figure}

The Data Processor links each telescope with the Gondola system..
It is a complex system which allows to control, to configure, to monitor and to operate each telescope during the commissioning phase, the test campaigns and during the flight.
In what follows we describe the DP architecture designed for the Fluorescence telescope, but the one foreseen for the Cherenkov telescope is very similar.
The DP of EUSO-SPB2 telescopes is an evolution of the one developed for EUSO-SPB1 \cite{scotti}
%The two telescopes use different sensors for the focal surface and different front-end electronics due to the characteristics of the signal they have to study. However, from point of view of the control, configuration, monitoring, data management as well as definition of the operative mode the two instruments are very similar.It is a complex system which allows to control and monitor the telescopes, during the commissioning, the test campaigns as well as in flight.
A block diagram of the DP is shown in Figure  \ref{fig:block_diag1}.
The main interface of the DP with the focal surface electronics is realized

 by the Clock Board (CLKb) which is a follow-up design of the board developed for the previous  EUSO missions \cite{scotti1}. The CLKb provides the time synchronization of the whole telescope,  as well as the tagging of an event with arrival time and payload position acquired by the two GPS receivers.

The Central Processor Unit (CPU) module is 
is an embedded computer with a CAN bus controller and hard disks. It is 
the interface of the DP with the NASA telemetry system and with all other equipment present on board. The data acquisition and the control of the focal surface electronics is performed through an Ethernet connection (Fig. \ref{fig:block_diag1} blue arrows), CAN bus (green bus) is used to monitor the voltages, the currents as well as the temperatures on the Zynq boards of the focal surface. CAN bus is also used to control and acquire data of the ancillary devices present on the telescope (AMON, EMON, Health LED, etc).
The capability to perform ordered power-ON and power-OFF sequences is provided by a Solid State Power Module (SSPM). The SSPM is controlled by the CPU via the CAN bus.

The House Keeping (HK) board reads out the temperature sensors and controls of the heaters.

The designed architecture allows the spare CPU to take over the full control of the telescope in case of failure of the main CPU.

The DP modules are hosted, together with the low voltage power supplies modules, in the DP mechanical sub-rack (Eurocard double size sub-rack).

\subsection{GPS receivers}
The two GPS receivers, part of the DP system, work in parallel for redundancy.
Each receiver provides timing signals to the CLKb board and exchanges data and commands through a serial communication port.
The time synchronization is ensured by the 1 Pulse Per Second (1 PPS) output of the receiver which are also sent to the CLKb-Board;
The selected receivers (Trimble bx992) are based on  336 Channels chips for multi-constellation GNSS support. To comply with International Traffic in Arms Regulations (ITAR) restrictions, there is a limit in velocity (515 $m/s$ above 18 km altitude). The working temperature range is -40$^\circ$C + 85$^\circ$C. 

\subsection{The Clock board}
The Clock-Board  hosts the interfaces with the GPS receivers. It 
%tags the events with their arrival time (UTC) and payload position.  
associates the telescope's position and velocity, UTC time, and satellite tracking status information measured by the GPS receivers  to each acquired event. 
The Clock-Board receives  first level trigger signals from the Zynq boards and provides  Veto and Busy signals to properly synchronize the data acquisition process.  
The Clock-Board  can generate  trigger signals of fixed frequency, on a CPU command or in coincidence with the 1 PPS signal of the GPS. 
It also measures the operating time  and dead time of the instrument and provides signals for time synchronization of the event. Most of the functions of the CLK Board will be implemented in a FPGA Xilinx  Zynq  XC7Z020 chip.

\subsection{CPU and Data Storage}
The CPU module is a low power PCI Express embedded controller. It is based on an
Core i7 3517UE 1.7 GHz Dual-Core processor with a DDR3 memory interface operating at up to 800 MT/s. A Serial-ATA (SATA) controller provides a 6 Gbit/sec connection to the hard drives. The CPU  collects data from the instruments through a Gigabit LAN port. 
The data are stored on-board using two 2 TB Solid-State Drives (SSD) operating in fault-tolerant RAID-1 configuration (Redundant Array of Independent Disks). A dual channel CAN bus interface is hosted on PCI/104-Express embedded controller. This High-speed CAN bus is used to monitor the house-keeping parameters of the focal surface electronics and to control the SSPM and the other ancillary devices.
The working temperature range of the CPU and Data Storage hard disk is -40$^\circ$C + 85$^\circ$C. 
A spare CPU will be present on board for redundancy.

\subsection{HK system}
The House-Keeping system collects telemetry information from the sub-systems of the instrument in slow control mode. It is responsible for monitoring temperatures and controlling the heaters in case of "cold" power ON. It uses a Ethernet link to transmit telemetry data and to receive command from the CPU.
The HK is implemented around a LabJack T7 module.  

\subsection{Solid State Power Module}
The 16-Channel Solid State Power Controller and Power Distribution Unit will be used  to switch battery power on/off  for
all connected consumers. This unit has a higher reliability and a greater system protection compared to  mechanical switches, breakers, and relays which improves the mission's safety and longevity. The SSPM unit will be controlled by the CPU via the CAN bus.

\section{The DP control and acquisition software}
\subsection{The DP control software}
The DP plays the fundamental role of being the interface between the detectorand the telemetry
blocks and the end-users. For this reasons, the control software will be designed, taking into account the experience of EUSO SPB  \cite{fornaro}. It will be flexible enough to be used in every phase of the detector commissioning and flight. It has to guarantee the access to the data, control and monitor the detector, as well as the storing of all
relevant data, i.e. scientific, housekeeping and logging ones. Moreover,  the design enables full user control via network, telemetry or console interfaces  as well as a quick and comprehensive user configuration. The modular design of control software allows different DP configurations during test, commissioning and flight phases and will  handle any potential block failure during the flight. Least but not last, the control software should permit development, test and detector control from different geographical location.

During the flight the status of the apparatus will be continuously checked  to distinguish condition to start the measurement (dark time) or stop the measurement (day time).
Four acquisition modes are defined:
\begin{itemize}
	\item Dark time: data taking and storing of events,
	\item Day time: data compression and transmission of data to ground
	\item Day to dark transition: check the light before switching the instruments all ON
	\item Night to day transition: switch OFF the HV
\end{itemize}
The acquisition modes will be selectable from ground. 
\subsection{The DP acquisition software}
The data acquisition software is the application used for science data acquisition from the CLK board and Zynq boards. 
In order to do that, the acquisition software performs many other operations: the initialization and configuring of the subsystems, monitoring the connections with the CPU, verifying the behaviour of some subsystem parts and  responding to external commands from the user or from control software, etc. It can also be instructed to perform many types of acquisitions, among them: recording of external events (cosmic rays or CLKB/GPS pulses), providing internal simulated events (CPU triggering). 
\section{Conclusion}
The general architecture of the Data Processor for the EUSO-SPB2 mission telescopes has been designed. The architecture foresees significant improvements with respect to the one realized for the previous missions. Most of the components of the DP have been selected (COTS devices) and others are in advanced stage of design. The integration and the test of the DP system is planned for the beginning of 2020. The integration with the focal surface electronics and the thermo-vacuum test of the two systems is planned for the 2nd quarter 2020.   
\section{Acknowledgements}

We would like to thank the staff at our home institutions for their strong and undivided support all along this project.

This work was partially supported by the Italian Space Agency through the ASI INFN agreement n. 2017-8-H.0.


\begin{thebibliography}{99}

\bibitem{Bertaina}
M. Bertaina et al. for the JEM-EUSO Collaboration, "Search for Ultra-High Energy Cosmic Rays from Space"
\emph{Proceedings of the 36$^{th}$ ICRC}, PoS(ICRC2019)192.
\bibitem{Adams1}
J. H. Adams et al, "White paper on EUSO-SPB2"
\emph{arXiv:1703.04513v1} (2017).
\bibitem{scotti0}
V. Scotti, G. Osteria, for the JEM-EUSO Collaboration, 
"The EUSO-SPB mission", 
\emph{DOI: https://doi.org/10.22323/1.314.0024} (2018).
\bibitem{poemma}
A.Olinto et al,
\emph{arXiv:1708.07599v1 [astro-ph.IM]} (2017).
\bibitem{lawrence}
L. Wiencke et al. for the JEM-EUSO Collaboration, "The Extreme Universe Space Observatory on a Super-Pressure Balloon II Mission"
\emph{Proceedings of the 36$^{th}$ ICRC}, PoS(ICRC2019)466.
\bibitem{trigger}
A. Belov, M. Bertaina, F. Capel,et al., "The integration and testing of the Mini-EUSO multi-level trigger system",
\emph{Advances in Space Research} V. 62 2966–2976 (2018).
\bibitem{otte}
N. Otte  et al. for the JEM-EUSO Collaboration, "Development of a Cherenkov Telescope for the Detection of Ultra-High Energy Neutrinos with EUSO-SPB2 and POEMMA"
\emph{Proceedings of the 36$^{th}$ ICRC}, PoS(ICRC2019)977.
\bibitem{amon}
S: Mackovjak et al, "Airglow monitoring by one-pixel detector"
\emph{Nuclear Instruments and Methods in Physics Research},  A922 pp.150-156
doi: 10.1016/j.nima.2018.12.073, (2019).
\bibitem{scotti}
V. Scotti et al., "The Data Processor system of EUSO-SPB1"
\emph{Nuclear Instruments and Methods in Physics Research}, A916, pp. 94-101,
doi: 10.1016/j.nima.2018.10.207, (2019).
\bibitem{scotti1}
V. Scotti and G. Osteria for the JEM-EUSO Collaboration, "The JEM-EUSO time synchronization system"
\emph{Nuclear Instruments and Methods in Physics Research}, A718 (2013), 248.

\bibitem{fornaro}
C. Fornaro, et al., "The onboard software of the EUSO-SPB pathfinder
experiment", 
\emph{Software - Practice and Experience} doi: 10.1002/spe.2655,(2019)
\end{thebibliography}
\end{document}